# Enhanced higher temperature (20-30 K) transport properties and irreversibility field in nano-$Dy_2O_3$ doped advanced internal Mg infiltration processed $MgB_2$ composites

G. Z. Li,[1,a)] M. D. Sumption,[1] M. A. Rindfleisch,[2] C. J. Thong,[2] M. J. Tomsic,[2] and E. W. Collings[1]

[1]*Department of Materials Science and Engineering, The Ohio State University, Columbus, Ohio 43210, USA*

[2]*Hyper Tech Research Incorporated, 539 Industrial Mile Road, Columbus, Ohio 43228, USA*

A series of $MgB_2$ superconducting composite strands co-doped with $Dy_2O_3$ and C were prepared via an advanced internal Mg infiltration (AIMI) route. The transport properties and $MgB_2$ layer growth were studied in terms of the $Dy_2O_3$ doping level, reaction temperature, and reaction time. Transport studies showed that both critical current densities, $J_c$s, and irreversibility fields, $B_{irr}$s, were increased with $Dy_2O_3$ doping. The highest layer $J_c$ was $1.35 \times 10^5$ A/cm$^2$ at 4.2 K, 10 T, 30 % higher than that of the best AIMI wires without $Dy_2O_3$ doping. The highest "non-barrier" $J_c$ reached $3.6 \times 10^4$ A/cm$^2$ at 4.2 K, 10 T, which was among the best results reported so far. The improvements were even more pronounced at higher temperatures where the field at which the layer $J_c$ reached $10^4$ A/cm$^2$ was pushed out by 0.9 T at 20 K, 1.2 T at 25 K, and 1.4 T at 30 K. While little or no enhancement in $B_{irr}$ was seen at 10 K and 15 K, the increases in $J_c$ at higher temperatures were consistent with observed increases in $B_{irr}$ of 17% at 20 K, 44% at 25 K, and 400% at 30 K. Also, there were some indications that the reaction and layer growth of $MgB_2$ was enhanced by $Dy_2O_3$ doping.

---
a) Author to whom correspondence should be addressed. Electronic mail: li.1423@osu.edu.

Porosity and connectivity have until recently been a significant limitation in the development of very high transport critical current density ($J_c$) MgB$_2$.[1] Conventional MgB$_2$ wires are produced either by the *in-situ* "powder in tube" (PIT) process, which reacts the Mg and B precursor mixture at typically 650-700 °C, or the *ex-situ* process where already reacted MgB$_2$ is filled into a tube and wire drawn followed by a final sintering at 900 °C. The $J_c$ of the *in-situ* best wire is typically larger than that of the *ex-situ*. Even so, due to volume shrinkage during the reaction, the *in-situ* PIT route generates a high quantity of dispersed voids, which limit the ultimate $J_c$. Notwithstanding, such PIT conductors[2,3] are well suited to various lower field applications even now, but the application niche of MgB$_2$ will be further expanded if $J_c$ can be increased.

The infiltration process was first used by Giunchi[4]. In this process a composite is formed in which a Mg rod is surrounded by B powders, all inside of an outer sheath. Upon heating, Mg liquid or vapor percolates into the B, and a shrinking core reaction forms MgB$_2$ locally. A nearly fully dense layer is formed, with the porosity accumulating in the central region previously filled by Mg. This method removes porosity from the superconducting layer and increases connectivity. Hur and his co-workers[5] pushed the $J_c$ of such "internal Mg diffusion" wires up substantially using SiC-doped fine B powders, instead of coarse B, as the precursor B. SiC additives provided a C source which increased the $B_{c2}$, and fine B powders led to increased grain boundary pinning leading to $J_c$s of $1.0 \times 10^5$ A/cm$^2$ at 4.2 K, 10 T.[6] Our group demonstrated increased layer $J_c$ with the use of pre-C doped nano-sized amorphous B powders.[7,8] Unfortunately, for the conductors above, the MgB$_2$ layer growth stopped shortly after the onset of heat treatment, leaving an unreacted B shell, and limiting the total current density $J_e$ ($I_c$/whole strand area) in the wire.[6,8] Recently, further advances have led to an understanding of the limiting layer thickness effect, allowing us to modify wire geometry/design, powder choice and preparation, as well as heat treatment conditions to minimize this effect.[9] Our advanced internal Mg infiltration process (AIMI) composite design accomplished the complete reaction of an infiltration based composite, resulting not only in a high layer $J_c$ but also an expanded $MgB_2$ fill factor, and a high $J_e$. The "non-barrier" $J_c$, defined by the critical current over the whole area inside the Nb barrier, reached 40 kA/cm$^2$ at 4.2 K, 10 T.[8,9]

Efforts are ongoing to further increase layer $J_c$s as well as to enlarge $MgB_2$ layer thickness. Recently, Mg particles have been added into B layers to shorten the Mg infiltration distance and increase the $MgB_2$ area with some success, but at high levels of Mg addition into the B layers voids are again generated.[10] To date, very few dopant species have been explored on infiltration-processed samples.[11] The few existing reports are about doping C or C-related chemicals.[8,12,13] A previous study conducted by our group shows that although C and carbides increase the layer $J_c$s remarkably, they suppress the $MgB_2$ layer formation at the same time.[8] Consequently, AIMI wires with high levels of C-doping (3-4%) are extremely difficult to react fully, limiting the $J_e$ of such a composite. It is of interest to see if a dopant can be found which can enhance the layer formation depth of infiltration process composites; dopants which can also increase $B_{c2}$ or pinning also remain relevant.

Here we explore the effects of $Dy_2O_3$ on AIMI processed $MgB_2$ wires. In previous studies, $Dy_2O_3$ was reported to enhance flux pinning and in-field $J_c$s for *in-situ* processed $MgB_2$.[14,15] In the present study, a small amount of $Dy_2O_3$ nanopowder has been co-doped with C in AIMI style composites. We see some indications that the reaction and layer growth of $MgB_2$ is enhanced. In addition, an improved layer $J_c$ of $1.35 \times 10^5$ A/cm$^2$ at 4.2 K, 10 T was attained for our best $Dy_2O_3$ and C co-doped AIMI wire. Most dramatically, however, the higher temperature transport properties and irreversibility field $B_{irr}$ were enhanced substantially. $Dy_2O_3$ shows itself to be the most successful dopant since C in its many forms for enhancing the properties of $MgB_2$, and does so in a regime where C is not effective – high temperatures.

A series of $MgB_2$ composite strands with a small amount of $Dy_2O_3$ addition was fabricated by Hyper Tech Research Inc. (HTR). The starting powders were pre-C-doped B from Specialty Materials Inc, and $Dy_2O_3$ (99.9%, 100 nm in size) from Sigma Aldrich. The B powders were mostly amorphous, 10-100 nm in size, with C doping level of 2 mol%.[16,17] They were mixed with 1 or 2 wt.% $Dy_2O_3$ nanopowders before the wire fabrication (the weight percentage, wt.%, is the weight of $Dy_2O_3$ over the weight of $Dy_2O_3$ and B mixture). The composite fabrication procedure followed that of our previous work on AIMI strands.[8,9] The strands were monofilamentary, with Monel outer sheaths, Nb chemical barriers, and a central region which consisted of a Mg rod surrounded by B powders. The final strand diameter was 0.52 mm. After fabrication, they were encapsulated under Ar, and reacted at 650 °C or 675 °C for 1-8 hours (ramp rate of 10 °C/min). Table 1 lists the $Dy_2O_3$ doping level, as well as the temperature and time of the reaction.

The microstructures and the $MgB_2$ layer formation were examined using an Olympus PME-3 optical microscope (OM), since as compared with scanning electron microscopy (SEM), OM images have better contrast between $MgB_2$ and B-rich layers. Transport $V$-$I$ measurements were performed on 50-mm-long segments cut from the wires and tested in transverse magnetic fields of up to 13 T in pool-boiling liquid He at 4.2 K. Subsequently, a 35-mm-long piece of sample D3 was mounted onto a variable-temperature insert (described in Ref 7), and $V$-$I$ measurements were carried out at transverse fields of 0-13 T and temperatures of 10-30 K. The electric field criterion $E_c$ for the transport measurements was 1 $\mu$V/cm and the sample gauge length in all the measurements was about 5 mm. The magnetic $J_c$s at temperature of 4.2-30 K and fields of 0-14 T was also characterized, using a Quantum Design Model 6000 Physical Property Measuring System (PPMS) as described elsewhere.[7]

Figure 1 shows the transverse cross-sectional OM images of the composites. As illustrated in the insert, the orange annulus is the $MgB_2$ layer and the dark annulus surrounding the $MgB_2$ is the partially reacted B-rich layer. $MgB_2$ layers of different thickness are formed depending on their respective $Dy_2O_3$ doping level and the reaction conditions. Samples A1-A4 (1 wt.% $Dy_2O_3$) are heat treated at 650 °C. Their $MgB_2$ layer thicknesses increase with reaction time. As shown in Figure 1, a higher heat treatment temperature (samples B1-B4) or a heavier $Dy_2O_3$ doping level (samples C1-C4) leads to a faster $MgB_2$ layer formation. Both groups of strands are entirely reacted within 8 hours. The $MgB_2$ layer growth rate is roughly doubled with either an increase from 1 wt.% to 2 wt.% $Dy_2O_3$ doping, or an increase of reaction temperature from 650°C to 675°C. The D-series composites (2 wt.% $Dy_2O_3$ and 675 °C) are fully reacted less than 4 hours.

The layer $J_c$ values at 4.2 K for the A and B series composites (1 wt.% $Dy_2O_3$) are shown in Figure 2 (a), while those for C and D series composites (2 wt.% $Dy_2O_3$) are shown in Figure 2(b). The 1 wt.% $Dy_2O_3$ doped AIMI composites high layer $J_c$s. The best sample B2 has the layer $J_c$ of $1.20 \times 10^5$ A/cm$^2$ at 10 T. However, C3 and D2 show the best layer $J_c$-B characteristics. The highest $J_c$ reaches $1.35 \times 10^5$ A/cm$^2$ at 10 T, which is 30% higher than that of the previously reported C-doped AIMI wires.[9] While the layer $J_c$ is useful for understanding intrinsic properties of the materials, if we wish to compare to PIT-processed $MgB_2$, we should consider the "non-barrier" $J_c$ defined using the total area inside of the Nb barrier.

Figure 3 presents the field dependences of "non-barrier" $J_c$s at 4.2 K. Even though sample C3 and D2 have the highest layer $J_c$s, their "non-barrier" $J_c$s are not the highest of the set due to the partial $MgB_2$ formation and the concomitant lower $MgB_2$ fill factor. The layer $J_c$s of fully reacted samples, C4 and D3, are slightly lower than those of C3 and D2 but their high $MgB_2$ fill factor gives them the highest "non-barrier" $J_c$s. The extrapolated "non-barrier" $J_c$ for sample D3 is $3.0 \times 10^5$ A/cm$^2$ at 5 T, and it reaches $3.6 \times 10^4$ A/cm$^2$ at 10 T. These preliminary results are very promising: the "non-barrier" $J_c$s are double those of state-of-the-art PIT wires.[3,18] They are also comparable to the best "non-barrier" $J_c$s of the C-doped AIMI wires.[8] It is believed that the "non-barrier" $J_c$s of $Dy_2O_3$ co-doped AIMI wires can be further enhanced after optimizations on $Dy_2O_3$ and C doping level and heat treatment conditions.

The enhancement of $J_c$s with $Dy_2O_3$ doping is even more clearly seen at higher temperatures. Figure 4(a) exhibits the layer $J_c$s of D3 at temperatures from 4.2-30 K and magnetic fields of 0-13 T. Compared with a C-doped infiltration-processed wire,[7] this sample shows enhanced layer $J_c$s at elevated temperatures. A layer $J_c$s of $10^4$ A/cm$^2$ is achieved at 10 K, 15 K, 20 K, 25 K, and 30 K at fields of 12.2 T, 9.6 T, 6.8 T, 4.0T, and 1.4 T, respectively. This represents an increase of 0.2 T, 0.5 T, 0.9 T, 1.2 T, and 1.4 T at those same temperatures over that of the non-$Dy_2O_3$ doped $MgB_2$ of Ref 7. Also as shown in Figure 4(b), the extrapolated "non-barrier" $J_c$ is $2.0 \times 10^4$ A/cm$^2$ at 20 K and 5 T, much higher than the $J_c$s of both PIT and C-doped infiltration-processed wires.[7] The irreversibility fields, $B_{irr}$s, are extracted from Figure 4(b), based on the criterion that $B_{irr}$ is equal to the field when the "non-barrier" $J_c$ reaches 1000 A/cm$^2$.[19] As compared in Table II, using same pre-C-doped boron, D3 shows higher $B_{irr}$s at all temperatures than the best C-doped PIT or C-doped infiltration-processed strands. This is important because, as is well known, C or C-related doping substantially increases $B_{irr}$ and $J_c$s of $MgB_2$ at 4.2 K, but is not effective at higher temperatures (20 K and above)[7,20] There is a clear improvement in both $J_c$ and $B_{irr}$ attributable to $Dy_2O_3$ doping, but calculations based on the magnetic $J_c$s of Figure 4 indicates that the maximum pinning strength $F_{p,max}$ for D3 is 54.0 GN/m$^3$ at 4.2 K and 11.4

GN/m$^3$ at 20 K, close to those of the C-doped-only infiltration-processed strand. Thus, it is clear that the increase in $J_c$ due to Dy$_2$O$_3$ doping is driven by increases in $B_{irr}$ (and thus presumably $B_{c2}$), rather than flux pinning,[14,15] for our AIMI composites.

Our study shows that MgB$_2$ infiltration and formation are enhanced by both increasing the reaction temperature from 650°C to 675°C, and with increases in Dy$_2$O$_3$ doping, which may offer us an option to fabricate infiltration-processed MgB$_2$ wires with both high layer $J_c$s and larger MgB$_2$ fill factors. In addition, the 4.2 K layer $J_c$ reached $1.35 \times 10^5$ A/cm$^2$ at 4.2 K, 10 T, 30 % higher than that of the AIMI wires without Dy$_2$O$_3$ doping. The highest "non-barrier" $J_c$ reached $3.6 \times 10^4$ A/cm$^2$ at 4.2 K, 10 T, which was among the best results reported so far. The increases are even more pronounced at higher temperatures where the field as which the $J_c$ reaches $10^4$ A/cm$^2$ is pushed out by 0.9 T at 20 K, 1.2 T at 25 K, and 1.4 T at 30 K. The $J_c$ increases seen at higher temperatures are consistent with observed increases in $B_{irr}$ of 17% at 20 K, 44% at 25 K, and 400% at 30 K. Dy$_2$O$_3$ shows itself to be the most successful dopant since C in its many forms for enhancing the properties of MgB$_2$, and does so in a regime where C is not effective – 20-30 K.

**ACKNOWLEDGMENTS**

This work was supported by NASA under SBIR contract NNX14CC11C.

**Table captions**

TABLE I. Doping levels and reaction times/temperatures of samples.

TABLE II. Comparison of $B_{irr}$ between $Dy_2O_3$/C co-doped AIMI sample D3 and best C-doped PIT and infiltration processed strands in Ref. 7. (unit: T)[a]

TABLE I. Doping levels and reaction times/temperatures of samples[a].

| Sample Name | $Dy_2O_3$ Doping Level, wt.% | Reaction Temperature, °C | Reaction Time, hr |
|---|---|---|---|
| A1 | 1 | 650 | 1 |
| A2 | 1 | 650 | 2 |
| A3 | 1 | 650 | 4 |
| A4 | 1 | 650 | 8 |
| B1 | 1 | 675 | 1 |
| B2 | 1 | 675 | 2 |
| B3 | 1 | 675 | 4 |
| B4 | 1 | 675 | 8 |
| C1 | 2 | 650 | 1 |
| C2 | 2 | 650 | 2 |
| C3 | 2 | 650 | 4 |
| C4 | 2 | 650 | 8 |
| D1 | 2 | 675 | 1 |
| D2 | 2 | 675 | 2 |
| D3 | 2 | 675 | 4 |
| D4 | 2 | 675 | 8 |

[a]All samples use pre-C-doped B, with 2 mol.% C with respect to B amount(see. Refs. 16, 17.)

TABLE II. Comparison of $B_{irr}$ between $Dy_2O_3$/C co-doped AIMI sample D3 and best C-doped PIT and infiltration processed strands in Ref. 7. (unit: T)[a]

| Temperature, K | D3 | C-doped AIMI | C-doped PIT |
|---|---|---|---|
| 10 | 14.7[b] | 14.4 | 12.0 |
| 15 | 11.4 | 10.9 | 9.7 |
| 20 | 8.2 | 7.0 | 7.0 |
| 25 | 4.9 | 3.4 | 4.0 |
| 30 | 2.0 | 0.4 | - |

[a]All three samples use the same pre-C-doped B from Specialty Materials Inc, with 2 mol.% C.[16,17] [b]14.7 T is extrapolated from Figure 4b.

**Figure captions**

Figure 1.  Optical microscopic (OM) images of $Dy_2O_3$/C co-doped AIMI composites with various reaction temperatures and times. The insert in the right corner shows an expanded view of sample C2 – the orange annulus is $MgB_2$ layer and the dark surrounding annulus is a B-rich layer.

Figure 2.  Field dependence of the layer $J_c$s for (a) 1 wt.% and (b) 2 wt.% $Dy_2O_3$/C co-doped samples measured at 4.2 K.

Figure 3.  Field dependence of the "non-barrier" $J_c$s of (a) 1 wt.% and (b) 2 wt.% $Dy_2O_3$/C co-doped samples measured at 4.2 K.

Figure 4.  Field dependence of (a) the layer $J_c$ and (b) the "non-barrier" $J_c$s of sample D3 at magnetic fields of 0-13 T and temperatures of 4.2-30 K (as shown in filled dotted curves). Two other sets of data are provided for comparison purposes: (i) Dashed lines: magnetic results at variable temperature and fields. (ii) Empty dotted curves: The results of a C-doped infiltration-processed $MgB_2$ composite[7] without $Dy_2O_3$. Both samples use same pre-C-doped boron.

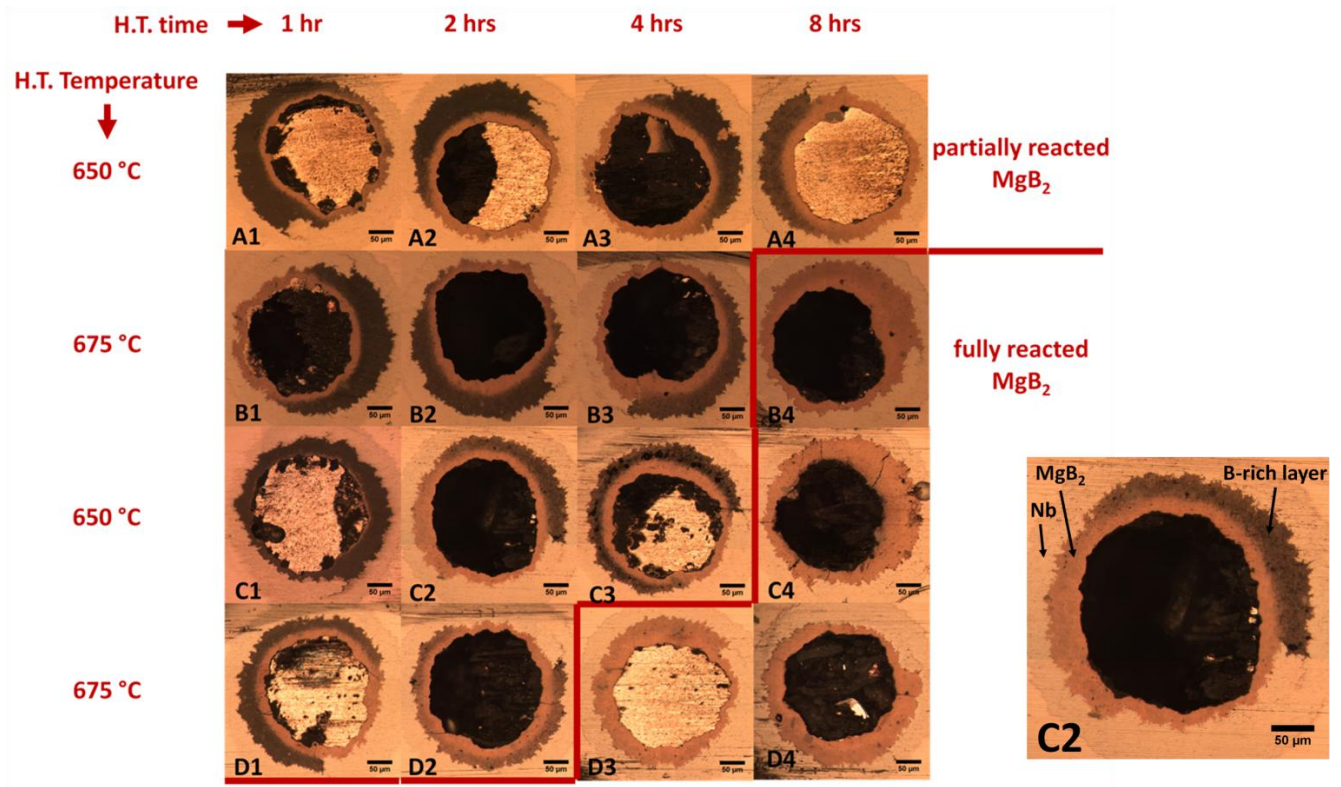

Figure 1. Optical microscopic (OM) images of $Dy_2O_3$/C co-doped AIMI composites with various reaction temperatures and times. The insert in the right corner shows an expanded view of sample C2 – the orange annulus is $MgB_2$ layer and the dark surrounding annulus is a B-rich layer.

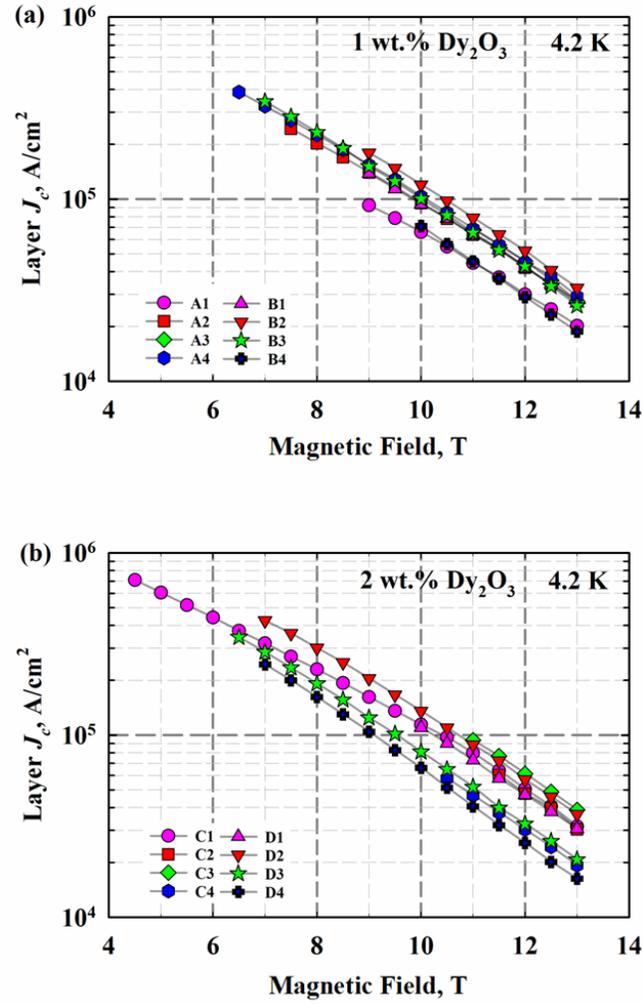

Figure 2. Field dependence of the layer $J_c$s for (a) 1 wt.% and (b) 2 wt.% $Dy_2O_3$/C co-doped samples measured at 4.2 K.

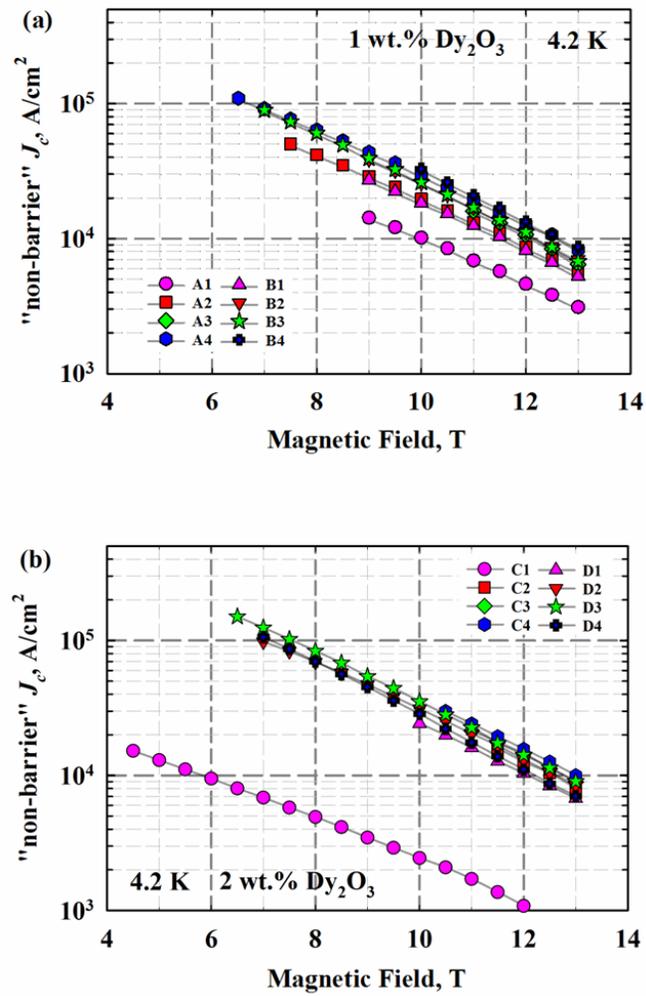

Figure 3. Field dependence of the "non-barrier" $J_c$s of (a) 1 wt.% and (b) 2 wt.% $Dy_2O_3$/C co-doped samples measured at 4.2 K.

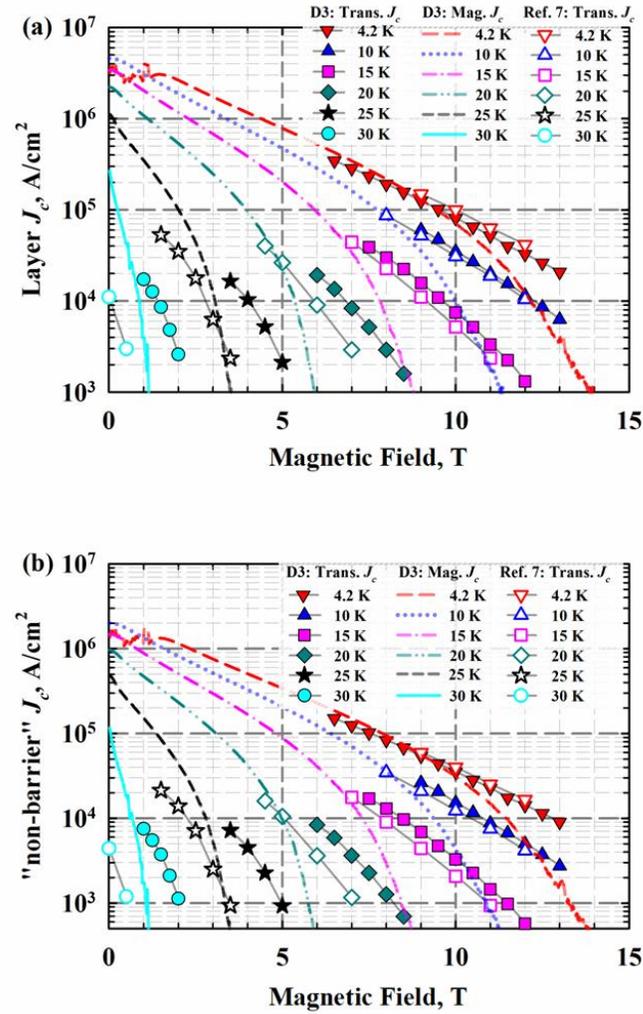

Figure 4. Field dependence of (a) the layer $J_c$ and (b) the "non-barrier" $J_c$s of sample D3 at magnetic fields of 0-13 T and temperatures of 4.2-30 K (as shown in filled dotted curves). Two other sets of data are provided for comparison purposes: (i) Dashed lines: magnetic results at variable temperature and fields. (ii) Empty dotted curves: The results of a C-doped infiltration-processed $MgB_2$ composite[7] without $Dy_2O_3$. Both samples use same pre-C-doped boron.